# Method for Generating Synthetic Data Combining Chest Radiography Images with Tabular Clinical Information Using Dual Generative Models.

Tomohiro Kikuchi (1,2), Shouhei Hanaoka (3), Takahiro Nakao (1), Tomomi Takenaga (3)
Yukihiro Nomura (1,4), Harushi Mori (2), Takeharu Yoshikawa (1)

1 Department of Computational Diagnostic Radiology and Preventive Medicine, The University of Tokyo Hospital, Tokyo, Japan.
2 Department of Radiology, Jichi Medical University, School of Medicine, Tochigi, Japan.
3 Department of Radiology, The University of Tokyo Hospital, Tokyo, Japan.
4 Center for Frontier Medical Engineering, Chiba University, Chiba, Japan.

The generation of synthetic medical records using Generative Adversarial Networks (GANs) is becoming crucial for addressing privacy concerns and facilitating data sharing in the medical domain. In this paper, we introduce a novel method to create synthetic hybrid medical records that combine both image and non-image data, utilizing an auto-encoding GAN (αGAN) and a conditional tabular GAN (CTGAN). Our methodology encompasses three primary steps: I) Dimensional reduction of images in a private dataset (pDS) using the pretrained encoder of the αGAN, followed by integration with the remaining non-image clinical data to form tabular representations; II) Training the CTGAN on the encoded pDS to produce a synthetic dataset (sDS) which amalgamates encoded image features with non-image clinical data; and III) Reconstructing synthetic images from the image features using the αGAN's pretrained decoder. We successfully generated synthetic records incorporating both Chest X-Rays (CXRs) and thirteen non-image clinical variables (comprising seven categorical and six numeric variables). To evaluate the efficacy of the sDS, we designed classification and regression tasks and compared the performance of models trained on pDS and sDS against the pDS test set. Remarkably, by leveraging five times the volume of sDS for training, we achieved classification and regression results that were comparable, if slightly inferior, to those obtained using the native pDS. Our method holds promise for publicly releasing synthetic datasets without undermining the potential for secondary data usage.

**INTRODUCTION**

Public data availability fosters reproducibility and new knowledge discovery through secondary analysis. Yet, sharing medical data is fraught with privacy issues due to its sensitive nature. A potential solution is synthetic data generation using generative models, which has been highlighted in recent times for its privacy-preserving qualities [1,2]. Through this, generative models, especially those based on Generative Adversarial Networks (GANs), can produce synthetic records resembling the original dataset, and many such methods have emerged in the medical domain [3–7].

However, the primary focus of existing synthesis methods has been either on image or tabular data, not both. Real-world data, especially in the medical sector, often comprises mixed types, like images and non-image clinical data. Such kind of hybrid data offers deeper insights into disease prognosis and personalized treatment. For instance, in the case of COVID-19, integrating chest X–ray radiographs (CXRs) with laboratory data can enhance prediction models [8].

Generating synthetic hybrid data that combines both images and non-image data has been challenging, primarily due to technical limitations like the curse of dimensionality and the complexity of inputting diverse data types. In this paper, we introduce a novel approach to generate synthetic hybrid records, combining both image and non-image data. Using two GANs (auto–encoding GAN ($\alpha$GAN) [9] and conditional tabular GAN (CTGAN) [10]), we created synthetic records containing both CXRs and non-image clinical data.

*Contribution of our Study*

1) Dimensionality reduction using a pre-trained encoder of auto-encoding GAN ($\alpha$GAN) [9] can simplify subsequent input to conditional tabular GAN (CTGAN) [10]. Notably, using the original hybrid dataset is not necessary for pretraining of $\alpha$GAN model because another image dataset can be utilized, eliminating the need for a large size of hybrid dataset as the source for synthesis.

2) With only 1,072 (training + validation) cases from original hybrid dataset, we successfully generated a synthetic hybrid database with intermediate–resolution (256 × 256 pixels) CXRs and thirteen non-image clinical data for each synthetic record.

*Related Works*

GAN, introduced by Goodfellow et al., has significantly contributed to the field of computer vision [11]. In the medical domain, GANs have been employed in a wide range of applications including modality transfer, super resolution, noise reduction, and synthetic medical record generation [4,12,13].

**Generation of CXRs:** For synthetic CXRs generation, several studies using nonconditional GANs have been proposed [5,14–16]. These studies used at least 5840 images with a minimum input image size of 244 × 244 pixels to train the networks. This implies that a large number of CXRs are required to train GANs with intermediate–to high–resolution images. In another study, synthetic CXR generation using conditional GAN was performed with 1124 CXRs used for training, but with a smaller input size of 112 × 112 pixels [17].

**Generation of medical tabular data:** The application of GAN to tabular data, such as TGAN, TableGAN, CTGAN, MedGAN, CopulaGAN, and CTAB–GAN, began a few years after its application to images [2,10,18–20]. In studies using tabular data generation for augmentation in the medical field, no architecture exists that can demonstrate superiority in all tasks; however, the performance of the CTGAN model seems to be stable [6,20,21].

## MATERIALS AND METHODS

*Overview of Our Study*

Figure 1 showed an overview of this study. We generated synthetic records in three steps: I) dimensional reduction of images in private dataset (pDS) using pretrained encoder of the αGAN model and merging with remaining non-image clinical data into tabular data; II) training the CTGAN model on the encoded pDS and generating synthetic dataset (sDS) which contained encoded image features and non-image data; and III) reconstruct the image features to synthetic CXRs using the pretrained decoder of the αGAN (Figure 1).

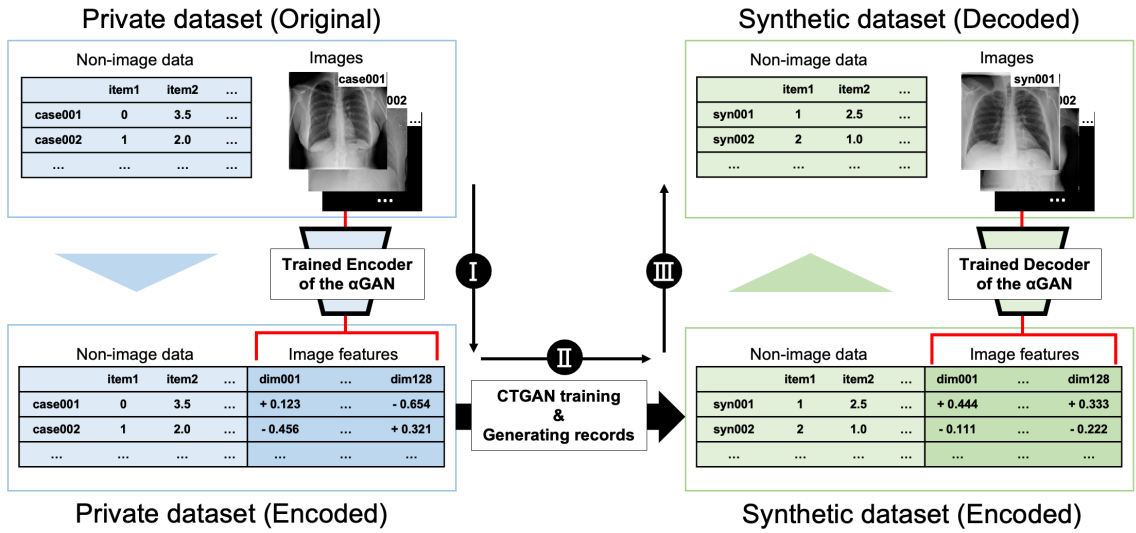

**Figure 1: Overview of the proposal method.**
Our approach consists of the following three steps: I) dimensional reduction of images in private dataset (pDS) using pretrained encoder of the αGAN model and merging with remaining non-image clinical data into tabular data; II) training the CTGAN model on the encoded pDS and generating synthetic dataset (sDS) which contains encoded image features and non-image clinical data; and III) reconstruct the image features into synthetic CXRs using the pretrained decoder of the αGAN.

*Datasets and preprocessing*

In this study, we assume that the Stony Brook University Covid–19 Positive Cases is the pDS. It is a publicly available hybrid dataset which contains CXRs and non-image clinical data [22,23]. This hybrid database comprises 1384 COVID–19 positive cases, featuring diverse imaging modalities and comprehensive non-image clinical tabular data. Originally, there were 130 variables per case in the non-image data. After removing variables with missing values greater than 5% and similar items, we ended up with 13 variables (Appendix 1). We randomly split the pDS into training, validation, and test datasets in a ratio of 6:2:2. The training and validation sets were used for training the CTGAN and constructing models for regression and classification tasks, while the test set was exclusively used for evaluating the performance of the regression and classification models. Table 1 presents the summary of the pDS used in the study.

For categorical variables, missing values were replaced with a new category representing the absence of data. For numerical variables, missing values were imputed using the mean values from the training and validation set. All images in the pDS were resized to 256 × 256 pixels before the implementation.

Table1. Summary of the contents of the dataset used as private dataset.

| Non-image items | Training set n = 800 | Validation set n = 272 | Test set n = 271 |
|---|---|---|---|
| **Categorical variables (number of patients)** | | | |
| Last status | | | |
| discharged | 692 | 237 | 235 |
| deceased | 108 | 35 | 36 |
| Age splits | | | |
| [18,59] | 428 | 168 | 149 |
| (59, 74] | 218 | 53 | 72 |
| (74, 90] | 154 | 51 | 50 |
| Gender concept name | | | |
| MALE | 439 | 152 | 153 |
| FEMALE | 343 | 110 | 114 |
| NA | 18 | 10 | 4 |
| Visit concept name | | | |
| Inpatient Visit | 588 | 204 | 203 |
| Outpatient Visit | 2 | 0 | 0 |
| Emergency Room Visit | 210 | 68 | 68 |
| Is ICU | | | |
| TRUE | 144 | 57 | 57 |
| FALSE | 656 | 215 | 214 |
| Was ventilated | | | |
| Yes | 120 | 47 | 46 |
| No | 680 | 225 | 225 |
| Acute kidney injury | | | |
| Yes | 147 | 51 | 50 |
| No | 653 | 221 | 221 |
| **Numeric variables (mean ± standard deviation)** | | | |
| Length of stay (day) | 10.0 ± 12.7 | 8.9 ± 9.6 | 10.1 ± 13.2 |
| (n. of missing value) | (0) | (0) | (0) |
| Oral temperature (°C) | 37.5 ± 0.9 | 37.5 ± 0.8 | 37.6 ± 0.9 |
| (n. of missing value) | (16) | (11) | (8) |
| Oxygen saturation (%) | 93.8 ± 5.5 | 93.5 ± 6.5 | 93.8 ± 5.6 |
| (n. of missing value) | (4) | (1) | (0) |
| Respiratory rate (/min) | 21.7 ± 7.5 | 22.2 ± 7.6 | 21.1 ± 6.7 |
| (n. of missing value) | (3) | (1) | (1) |
| Hate rate (/min) | 97.9 ± 19.4 | 100.6 ± 19.4 | 98.8 ± 22.0 |
| (n. of missing value) | (5) | (0) | (1) |
| Systolic blood pressure (mmHg) | 129.4 ± 23.1 | 128.7 ± 22.1 | 129.0 ± 22.4 |
| (n. of missing value) | (3) | (1) | (0) |

*This dataset is a subset of the Stony Brook University Covid-19 Positive Cases. The description for each variable is provided in Appendix 1. Please also refer to the information provided in the original dataset:
https://wiki.cancerimagingarchive.net/pages/viewpage.action?pageId=89096912

*Networks and Synthetic hybrid data generation*

We adopted αGAN as a network to compress/reconstruction the image dimensions. αGAN is a structure that combines variational auto–encoders (VAE) [9] with GAN and has been reported to produce sharper images than VAE while enabling the acquisition of latent variables that follow a Gaussian distribution similar to VAE [9,24,25]. In this study, the Radiological Society of North America (RSNA) Pneumonia Detection Challenge dataset (a publicly available CXRs dataset) [26] was used for αGAN model pretraining. We trained the αGAN to transition between CXRs of size 256 × 256 pixels and their corresponding 128-dimensional latent representations. Please refer to Appendix 2 for pre-training of αGAN.

We adopted CTGAN as the network for synthetic tabular data generation [10]. The CTGAN is a GAN–based model for modeling the distribution of tabular data. It performs normalization on each column of complex data distributions with respect to categorical variables, and trains the model using a conditional generator and discriminator. We used the standalone library of the CTGAN ver. 0.7.4 (https://github.com/sdv–dev/CTGAN). We trained the CTGAN model on encoded pDS excluding its test set and created 40,000 synthetic records as combined data containing image features and non-image data (encoded sDS). Image reconstruction can be performed by passing the synthetic image features in encoded sDS to the pretrained generator of αGAN model to obtain decoded pDS.

In addition, we prepared an unmatched dataset (uDS) as a benchmark for comparison with the proposed method. The uDS was also generated using the trained generator of the CTGAN; however, we split the generated data into tabular and image feature components, shuffled them, and merged them (Figure 2). Consequently, the uDS was formed with no correspondence between the tabular and image components. Similar to the sDS, 40,000 synthetic record instances were generated for the uDS.

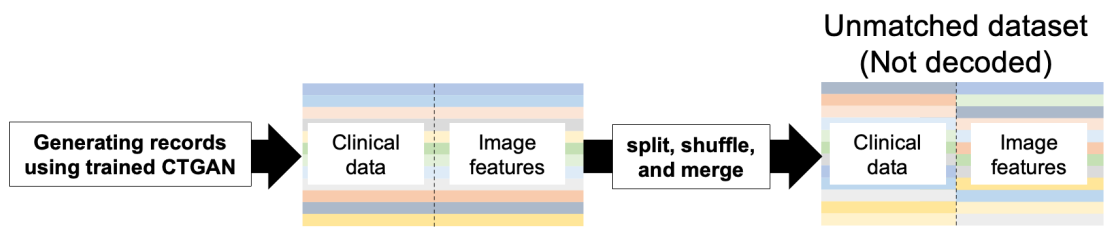

**Figure 2. Generating unmatched dataset for comparison.**
We split the output of the trained generator of the CTGAN model into non-image data and image features, shuffled their order, and merged them. We defined this database as an unmatched dataset (uDS).

*E. Evaluation methods*
1. **Visualization of similarity between datasets**
To visualize the similarities among the datasets, we employed t–stochastic neighbor embedding (SNE) plots [27,28]. We amalgamated 1072 cases from the encoded pDS (comprising both training and validation sets) and an equal number randomly sampled from the encoded sDS into a single data frame. Subsequently, we constructed a two-dimensional t-SNE plot to illustrate the distribution differences and visually evaluate the overlap between the two datasets. The t–SNE implementation was performed using the scikit-learn library ver. 1.3.0 (https://scikit-learn.org/stable/index.html).
2. **Usability**
To assess the performance of the decoded sDS, we implemented two classification tasks: "Last Status" and "Gender Concept Name", and two regression tasks: "Oral Temperature" and "Oxygen Saturation" (Please refer to Appendix 1 for descriptions of these variables). For evaluations involving non-image clinical data, we employed LightGBM version 3.3.4 (https://github.com/microsoft/LightGBM), whereas for evaluations centered around CXRs, ResNet50 [29] from Torchvision ver. 0.14.1 (https://pytorch.org/vision/stable/index.html) was our choice (See Appendix 3 for training parameters for LightGBM and ResNet50). First, the performance of training with pDS was measured. Then, we evaluated performance with sDS. During the training phase of sDS, the number of synthetic records was evaluated in two scenarios: equivalent to pDS and five times the size of pDS. This process was repeated five times in each scenario. In all experimental conditions, the test set from pDS was employed for evaluation. For the tasks with CXRs, assessments with uDS were also conducted. For the classification tasks, we evaluated performance using the Area Under the Receiver Operating Characteristic (AUROC), while for the regression tasks, we employed the Mean Absolute Error (MAE) as the evaluation metric.

## RESULTS

Figure 3 shows four examples of decoded sDS. These hybrid records are considered to indicate the high quality and variety achieved by the proposed approach. We present the first twelve CXRs of the decoded sDS in Appendix 4 to demonstrate the variety of images generated.

|  | example 1 | example 2 | example 3 | example 4 |
|---|---|---|---|---|
| Decoded image | 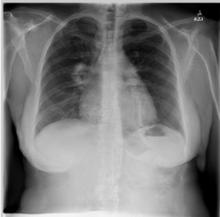 | 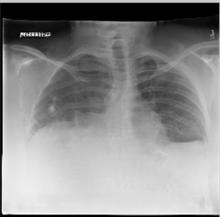 | 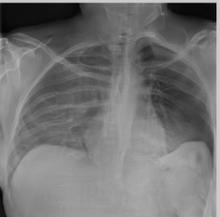 | 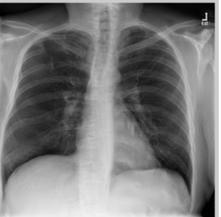 |
| Last Status | discharged | discharged | deceased | discharged |
| Gender Concept Name | FEMALE | FEMALE | MALE | MALE |
| Oral Temperature | 36.9 | 39.2 | 37.7 | 38.7 |
| Oxygen Saturation | 95 | 84 | 93 | 97 |

**Figure 3. Examples from decoded synthetic dataset.**
Examples of generated synthetic chest X-ray radiographs, accompanied by the non-image clinical data used in classification and regression tasks.

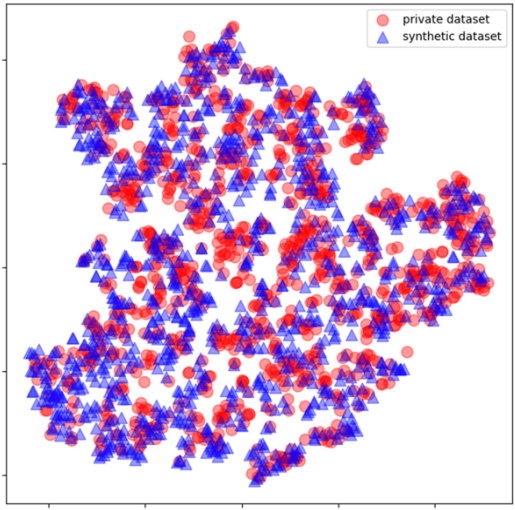

**Figure 4. Two–dimensional plot based on t–Stochastic Neighbor Embedding.**
Records from the encoded pDS are shown in red, those from the encoded sDS in blue.
pDS: private dataset, sDS: synthetic dataset

Figure 4 presents a t-SNE plot using the encoded pDS and encoded sDS. The overlap between the two is good.

Table 2 shows the results of the classification task. non-image data performed better than image data when targeting "Last Status" (AUROC: 0.65 for pDS images compared to 0.96 for pDS non-image data). When sDS was used, the results were similar to or slightly inferior to those of pDS. When targeting "Gender Concept Names", results were better with image data than with non-image data (AUROC: 0.79 for pDS images compared to 0.55 for pDS non-image data). The results with sDS were similar to or slightly inferior to those with pDS. When uDS is used for training, the 95% confidence intervals for AUROC includes 0.5, indicating that the inference is random.

**Table 2. Results of classification tasks**

| Target variable | Used dataset for training (n. of records) | AUROC (95% Confidence Intervals) | |
| --- | --- | --- | --- |
| | | Non-image data | Images |
| Last Status (discharge, deceased) | Private dataset (pDS) (n = 1072) | 0.96 | 0.65 |
| | Synthetic dataset (n = same as pDS) | 0.90 (0.89 – 0.92) | 0.59 (0.54 – 0.64) |
| | Synthetic dataset (n = 5 times of pDS) | 0.91 (0.90 – 0.93) | 0.65 (0.60 – 0.69) |
| | Unmatched dataset (n = same as pDS) | – | 0.51 (0.43 – 0.59) |
| Gender Concept Name* (MALE, FEMALE) | Private dataset (pDS) (n = 1044) | 0.55 | 0.79 |
| | Synthetic dataset (n = same as pDS) | 0.53 (0.46 – 0.59) | 0.59 (0.53 – 0.66) |
| | Synthetic dataset (n = 5 times of pDS) | 0.56 (0.52 – 0.60) | 0.72 (0.66 – 0.77) |
| | Unmatched dataset (n = same as pDS) | – | 0.52 (0.43 – 0.62) |

* Missing value codes are excluded from learning and testing

Table 3 shows the results of the regression task. Regardless of whether the target was "Oral Temperature" or "Oxygen Saturation", the training using non-image data appeared to yield superior results (when employing the pDS, the MAE was 0.61 for 'Oral Temperature' with non-image data and 0.65 with images. For 'Oxygen Saturation', the MAE was 3.1 with non-image data and 3.5 with images). The results with sDS were similar to or slightly inferior to those with pDS. The results with uDS were clearly worse than those with pDS or sDS.

**Table 3. Results of regression tasks**

| Target variable | Used dataset for training (n. of records) | MAE (95% Confidence Intervals) | |
| --- | --- | --- | --- |
| | | Non-image data | Images |
| Oral Temperature (°C) | Private dataset (pDS) (n = 1072) | 0.61 | 0.65 |
| | Synthetic dataset (n = same as pDS) | 0.63 (0.63 – 0.64) | 0.65 (0.63 – 0.66) |
| | Synthetic dataset (n = 5 times of pDS) | 0.62 (0.61 – 0.63) | 0.64 (0.64 – 0.65) |
| | Unmatched dataset (n = same as pDS) | – | 0.81 (0.78 – 0.84) |
| Oxygen Saturation (%) | Private dataset (pDS) (n = 1072) | 3.1 | 3.5 |
| | Synthetic dataset (n = same as pDS) | 3.4 (3.3 – 3.5) | 3.6 (3.5 – 3.7) |
| | Synthetic dataset (n = 5 times of pDS) | 3.3 (3.2 – 3.3) | 3.5 (3.4 – 3.5) |
| | Unmatched dataset (n = same as pDS) | – | 4.7 (4.4 – 4.9) |

## V. DISCUSSION

In this study, we proposed a novel approach by training an architecture capable of dimensionality reduction (in this study, αGAN) on a large open dataset, compressing the dimensions of pDS images through this model and passing them through a tabular–compatible GAN (in this study, CTGAN). We successfully generated synthetic records of hybrid CXR and non-image clinical data while maintaining the correspondence between them. The records generated using our method are diverse and can contribute to various secondary analyses. To the best of our knowledge, this is the first study to simultaneously generate hybrid CXR and medical tabular data. GANs are used to generate virtual patient cases for two distinct purposes: data augmentation and data sharing [2,4,5,13,18]. For augmentation, generating a subset of the database using the information required to solve a predetermined task is sufficient. However, for data sharing, synthetic records that mimic all variables and their characteristics in the original database maximize the potential for secondary analyses. However, training GANs on a database containing high–dimensional data is challenging if the database is not large, and the datasets collected by medical researchers are generally not large. To address this difficulty, we devised a method to effectively reduce the dimensionality of images by image encoder of αGAN model which was pretrained on another image dataset. We then combined the latent vectors with non-image data and trained the CTGAN model on the joint data (encoded pDS). Consequently, we successfully created synthetic records containing intermediate–resolution CXRs and a large amount of tabular data from 1072 training–validation datasets, which is a relatively small number of records for training GANs for intermediate–resolution image generation [5,14–16].

From Figures 3-4, we can infer that mode collapse, a frequent issue with GANs, does not occur. The overlap on the t-SNE plot between pDS and sDS is notable, suggesting that sDS effectively emulates the pDS. However, as observed in Appendix 3, there are broken images at low frequencies, indicating that not every record in the sDS is flawless.

We also conducted classification tasks, specifically "Last Status" and "Gender Concept Came", as well as regression tasks for "Oral Temperature" and "Oxygen Saturation". For the "Last Status" classification, non-image data provided a higher predictive accuracy than using images, with an AUROC of 0.91 for non-image data and 0.65 for image data in pDS (Table 2). In contrast, for 'Gender Concept Name', images yielded a higher predictive accuracy than non-image data, registering an AUROC of 0.55 for non-image and 0.79 for image data in pDS. When using sDS for classification, the results were comparable to pDS for "Last Status" but slightly inferior for "Gender Concept Name". In regression tasks, both "Oral Temperature" and "Oxygen Saturation" generally showed slightly better accuracy with non-image data, though image data also show similar MAE values (Table 3). The regression results with sDS were slightly inferior to pDS for 'Oral Temperature' but on par for 'Oxygen Saturation'. Generally, when training with sDS using the same number of records as pDS, the performance often lagged and by increasing the number of sDS records to five times that of pDS, the accuracy substantially approached that of the original sDS. This could be attributed to the presence of imperfect data, as also indicated in Appendix 3. However, for both classification and regression, the accuracy of sDS was consistently better than that of uDS, suggesting that we have successfully generated meaningful images corresponding to table data. Depending on the task, either non-image or image data may have better predictive power and may not be predictable a priori. A significant advantage of our method is its capability to produce fictional datasets that retain all data elements of the original dataset, thereby offering extremely high potential for secondary use.

Sharing medical data is valuable and important for ensuring the reproducibility of research and obtaining new insights through secondary analysis. However, the public release of the entire database is not easy because of privacy concerns. Our method can generate synthetic records for both CXRs and non-image clinical data simultaneously and is considered to have high secondary usability while considering privacy. While our method depends on the availability of a large-scale open dataset for the pretraining of the αGAN model, we believe it is a significant achievement to eliminate the necessity for a vast hybrid dataset as a synthesis source. This is particularly important because datasets collected by one or a few medical researchers, encompassing both images and tabular data, are generally limited in size. Thus, our approach proves valuable even in scenarios with constrained dataset sizes.

A significant potential competitor to our research is the language learning model (LLMs). Recent advances in LLMs have enabled these systems to handle multimodal inputs [30,31]. This capacity to manage and manipulate abstract information within the latent space overlaps with the objectives of our current methodology to a certain extent. However, our approach distinctly focuses on the generation of synthesized data that closely aligns with the characteristics of the original data. This targeted outcome distinguishes the proposed method and demonstrates its unique value. Although both LLMs and our technique share a common underpinning in the form of advanced learning mechanisms, their roles remain divergent because of their differentiated goals. As the field continues to evolve, we anticipate that the functionalities of these two methodologies will retain their individuality and serve complementary roles in various applications.

*future works*
1. **Application to data augmentation.** Although this study focuses on the publication of a dataset, the proposed framework can also be used for data augmentation.
2. **Extension to 3–Dimensional images.** CXRs were used in this study. If this method could be extended to 3–Dimensional images such as CT images, it could demonstrate further usefulness.
3. **Introducing a more rigorous privacy concept.** In privacy research, GAN outputs may contain information from the original data. The concept of differential privacy (DP), which was developed in recent years, addresses this issue [32]. DP can be applied to the proposed method to ensure robust privacy.

*limitations*

First, in our implementation, we used αGAN and CTGAN; however, it was possible to replace them with other networks. We did not evaluate the way the performance changes when the architectures are replaced. We referred to previous studies on αGAN and CTGAN for medical records and selected them for the proposed method. Second, the minimum number of records in pDS and the maximum number of columns in pDS to generate meaningful sDS are unknown. Our experimental results serve as indicators for future research.

*conclusion*

Using αGAN and CTGAN models, we generated synthetic hybrid records consisting of intermediate–resolution CXRs and tabular data with 70 variables. Although we assumed the availability of another large–scale image database, the proposed method enables the creation of highly reusable synthetic hybrid data without releasing the original hybrid database.

# SUPPLEMENTARY MATERIALS

**Appendix 1: Description of the variables used in our experiment.**

| Variable name | Possible Values | Variable Description* |
|---|---|---|
| Last Status | deceased, discharged | For the selected visit the status of the patient |
| Age Splits | [18,59], (59, 74], (74, 90] | Age intervals (in years) at time of admission |
| Gender Concept Name | FEMALE, MALE | Documented gender in the EHR (Electronic Health Record) |
| Visit Concept Name | Inpatient Visit<br>Outpatient Visit<br>Emergency Room Visit | For the selected visit the type of the visit |
| Is ICU | True, False | Patient admitted to the ICU based on documented room charges |
| Was Ventilated | Yes, No | The patient had invasive ventilation |
| Acute Kidney Injury | Yes, No | Had an increase in serum creatinine of 0.3 mg/dL within 48 hours |
| Length of Stay | Numeric value (days) | Number of calendar days in the facility |
| Oral Temperature | Numeric value (°C) | |
| Oxygen Saturation | Numeric value (%) | Oxygen saturation in Arterial blood by Pulse oximetry |
| Respiratory Rate | Numeric value (/min) | |
| Hate Rate | Numeric value (/min) | |
| Systolic Blood Pressure | Numeric value (mmHg) | |

\* The descriptions are extracted from:
https://wiki.cancerimagingarchive.net/pages/viewpage.action?pageId=89096912

**Appendix 2(a). Training of Auto-Encoding GAN with open dataset**

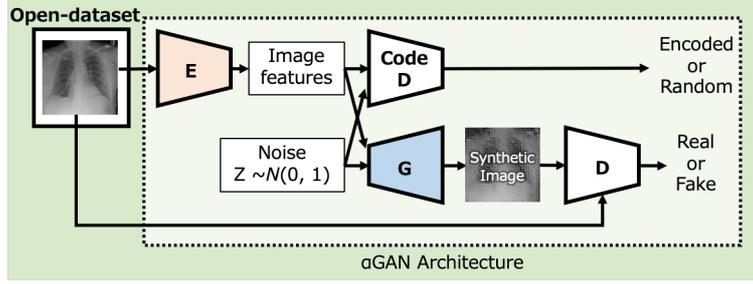

As a pre-training step to applying αGAN to the private dataset, the model is trained on an open database. In this study, the Radiological Society of North America (RSNA) Pneumonia Detection Challenge dataset was used for pretraining [1]. Encoder (E): Takes real images and encodes them into latent codes, representing a compressed representation of the input images. Generator (G): Uses the latent codes generated by E to produce fake (generated) images that resemble the real images. Discriminator (D): Determines whether an image is real (from the dataset) or fake (generated by G). Code Discriminator (Code D): Differentiates between real latent codes (sampled from the Gaussian distribution) and fake latent codes (generated by E). The trained E is used for dimension reduction (Figure 1- I), while the trained G is reused for image reconstruction (Figure 1-III).

**Appendix 2(b). The Auto-Encoding GAN architecture in our study.**

Generator

| Layer | Activation | Output shape |
|---|---|---|
| (Latent vector) | | 128 |
| Linear | LReLU | $4 \times 4 \times 512$ |
| Upsampling | | $8 \times 8 \times 512$ |
| Convolution 3×3 | LReLU | $8 \times 8 \times 256$ |
| Upsampling | | $16 \times 16 \times 256$ |
| Convolution 3×3 | LReLU | $16 \times 16 \times 128$ |
| Upsampling | | $32 \times 32 \times 128$ |
| Convolution 3×3 | LReLU | $32 \times 32 \times 64$ |
| Upsampling | | $64 \times 64 \times 64$ |
| Convolution 3×3 | LReLU | $64 \times 64 \times 32$ |
| Upsampling | | $128 \times 128 \times 32$ |
| Convolution 3×3 | LReLU | $128 \times 128 \times 16$ |
| Upsampling | | $256 \times 256 \times 16$ |
| Convolution 3×3 | LReLU | $256 \times 256 \times 8$ |
| Convolution 1×1 | Tanh | $256 \times 256 \times 1$ |

Encoder

| Layer | Activation | Output shape |
|---|---|---|
| (Input Image) | | $256 \times 256 \times 1$ |
| Convolution 1×1 | LReLU | $256 \times 256 \times 8$ |
| Convolution 3×3 | LReLU | $256 \times 256 \times 16$ |
| Downsampling | | $128 \times 128 \times 16$ |
| Convolution 3×3 | LReLU | $128 \times 128 \times 32$ |
| Downsampling | | $64 \times 64 \times 32$ |
| Convolution 3×3 | LReLU | $64 \times 64 \times 64$ |
| Downsampling | | $32 \times 32 \times 64$ |
| Convolution 3×3 | LReLU | $32 \times 32 \times 128$ |
| Downsampling | | $16 \times 16 \times 128$ |
| Convolution 3×3 | LReLU | $16 \times 16 \times 256$ |
| Downsampling | | $8 \times 8 \times 256$ |
| Convolution 3×3 | LReLU | $8 \times 8 \times 512$ |
| Downsampling | | $4 \times 4 \times 512$ |
| Linear | | 128 |

Discriminator

| Layer | Activation | Output shape |
|---|---|---|
| (Input Image) | | $256 \times 256 \times 1$ |
| Convolution 1×1 | LReLU | $256 \times 256 \times 8$ |
| Convolution 3×3 | LReLU | $256 \times 256 \times 16$ |
| Downsampling | | $128 \times 128 \times 16$ |
| Convolution 3×3 | LReLU | $128 \times 128 \times 32$ |
| Downsampling | | $64 \times 64 \times 32$ |
| Convolution 3×3 | LReLU | $64 \times 64 \times 64$ |
| Downsampling | | $32 \times 32 \times 64$ |
| Convolution 3×3 | LReLU | $32 \times 32 \times 128$ |
| Downsampling | | $16 \times 16 \times 128$ |
| Convolution 3×3 | LReLU | $16 \times 16 \times 256$ |
| Downsampling | | $8 \times 8 \times 256$ |
| Convolution 3×3 | LReLU | $8 \times 8 \times 512$ |
| Downsampling | | $4 \times 4 \times 512$ |
| Linear | | 1 |

Code Discriminator

| Layer | Activation | Output shape |
|---|---|---|
| (Latent vector) | | 128 |
| Linear | LReLU | 1500 |
| Linear | | 1 |

# Appendix 3. Parameters for model training

| | | |
|---|---|---|
| LightGBM | Classification | boosting_type: "goss", max_depth: 5, metric: "auc", num_boost_round: 1000, early_stopping_rounds: 100 |
| | Regression | boosting_type: "goss", max_depth: 5, metric: "mae", num_boost_round: 1000, early_stopping_rounds: 100 |
| ResNet50 | Classification | max epochs: 1000, early stopping duration: 20, batch size: 20, loss function: BCEWithLogitsLoss, optimizer: Adam, learning rate = 0.0001, data augmentation: Random Affine |
| | Regression | max epochs: 1000, early stopping duration: 20, batch size: 20, loss function: L1Loss, optimizer: Adam, learning rate = 0.0001, data augmentation: Random Affine |
| * Parameters not listed use the default values of the libraries in the study environment | | |

# Appendix 4: The first 12 CXRs of the synthetic dataset.

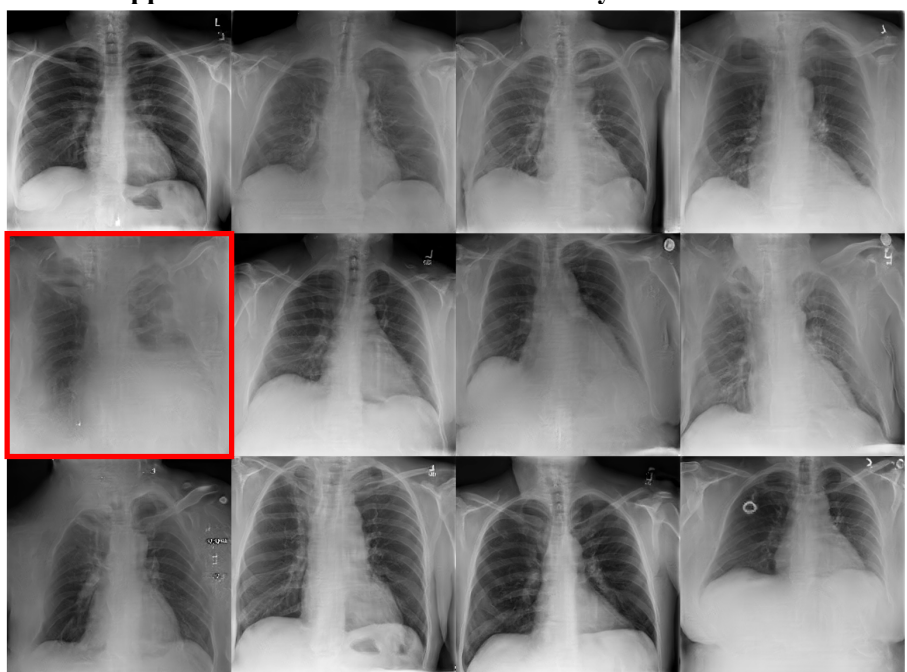

The image on the left of the middle row is slightly degraded (outlined in red), but the others are of a quality that allows them to be interpreted as CXRs.